\documentclass[final,english]{bullsrsl}[2022/06/15]

\usepackage[latin1]{inputenc}
\usepackage[T1]{fontenc}

\usepackage{natbib} 
\usepackage{graphicx}

\begin{document}
\title{SunPhot: Preparations for an upcoming quasar variability survey with the International Liquid Mirror Telescope}

\author[affil={1}, corresponding]{Ethen}{Sun}
\author[affil={2,3}]{Bhavya}{Ailawadhi}
\author[affil={4,5}]{Talat}{Akhunov}
\author[affil={6}]{Ermanno}{Borra}
\author[affil={2,7}]{Monalisa}{Dubey}
\author[affil={2,7}]{Naveen}{Dukiya}
\author[affil={1}]{Jiuyang}{Fu}
\author[affil={1}]{Baldeep}{Grewal}
\author[affil={1}]{Paul}{Hickson}
\author[affil={2}]{Brajesh}{Kumar}
\author[affil={2}]{Kuntal}{Misra}
\author[affil={2,3}]{Vibhore}{Negi}
\author[affil={2,8}]{Kumar}{Pranshu}
\author[affil={9}]{Jean}{Surdej}

\affiliation[1]{Department of Physics and Astronomy, University of British Columbia, 6224 Agricultural Road, Vancouver, BC V6T 1Z1, Canada}
\affiliation[2]{Aryabhatta Research Institute of Observational sciencES (ARIES), Manora Peak, Nainital, 263001, India}
\affiliation[3]{Department of Physics, Deen Dayal Upadhyaya Gorakhpur University, Gorakhpur, 273009, India}
\affiliation[4]{National University of Uzbekistan, Department of Astronomy and Astrophysics, 100174 Tashkent, Uzbekistan}
\affiliation[5]{Ulugh Beg Astronomical Institute of the Uzbek Academy of Sciences, Astronomicheskaya 33, 100052 Tashkent, Uzbekistan}
\affiliation[6]{Department of Physics, Universit\'{e} Laval, 2325, rue de l'Universit\'{e}, Qu\'{e}bec, G1V 0A6, Canada}
\affiliation[7]{Department of Applied Physics, Mahatma Jyotiba Phule Rohilkhand University, Bareilly, 243006, India}
\affiliation[8]{Department of Applied Optics and Photonics, University of Calcutta, Kolkata, 700106, India}
\affiliation[9]{Institute of Astrophysics and Geophysics, University of Li\`{e}ge, All\'{e}e du 6 Ao$\hat{\rm u}$t 19c, 4000 Li\`{e}ge, Belgium}

\correspondance{ethensun@astro.ubc.ca}
\date{\today}
\maketitle


%

\begin{abstract}
Recent research suggests a correlation between the variability and intrinsic brightness of quasars. If calibrated, this could lead to the use of quasars on the cosmic distance ladder, but this work is currently limited by lack of quasar light curve data with high cadence and precision. The Python photometric data pipeline SunPhot is being developed as part of preparations for an upcoming quasar variability survey with the International Liquid Mirror Telescope (ILMT). SunPhot uses aperture photometry to directly extract light curves for a catalogue of sources from calibrated ILMT images. SunPhot v.2.1 is operational, but the project is awaiting completion of ILMT commissioning.
\end{abstract}


\section{Introduction}
	The International Liquid Mirror Telescope (ILMT) is a new 4-metre liquid mirror telescope located at Devasthal Peak ($79^{\circ} 41' 07.1'', +29^{\circ} 21' 40.4''$, 2378m) \citep{Kumar_2018} in northern India. It is the only liquid mirror telescope currently in operation and uses a stabilized rotating container of mercury instead of a glass surface as a parabolic primary mirror. The cost of this type of mirror is much less but has the limitation that the telescope is fixed in a vertical orientation. Using time-delay integration, where electrons are shifted across the CCD at the sidereal rate, the ILMT obtains 102 second exposures of a 22-arcminute wide strip of sky.
	
	The ILMT saw first light in April 2022 and collected engineering data until the summer monsoon closure from June through early October. The climate and remoteness of the site have been obstacles in the commissioning, and the onsite team are continuing to work on the focus and mirror balance to improve image quality. In the meantime, a data pipeline for an ILMT quasar survey is being developed.
	
	Quasars are the most luminous objects in the known universe and emit extreme amounts of energy comparable to that of entire galaxy clusters. Recent research suggests a linear correlation between the absolute magnitude of quasars and the characteristic rate of variation in their light curves \citep{Solomon_2022}. The characteristic variational rate is the most common slope found in the light curve, using linear fits to a moving window and corrected for cosmological time dilation. Establishing and calibrating such a link could extend the cosmic distance ladder to as far as quasars can be observed, and contribute to cosmological measurements such as $H_0$. 

	Studies of quasar light curves are limited by available data, of which there are two types. Gravitational microlensing experiments observe dense star fields in the Magellanic Clouds and Galactic centre as frequently as multiple times a night. They have produced several hundred high-cadence, decade-long light curves for quasars found in these fields. These are of the highest quality, but are very limited in quantity and in sky coverage \citep{Geha_2003, Kozlowski_2013}. The second type is extracted from long-term surveys and transient searches, which produce hundreds of thousands of lightcurves using facilities like the Catalina Real-time Transient Survey (CRTS) and Zwicky Transient Facility (ZTF). These achieve limiting magnitudes of 20-21 \citep{Djorgovski_2011, Graham_2019} using moderate ($\sim$1.5 m) sized telescopes, but their sampling intervals range from as little as a few days to sporadic.

\section{Preparations for the upcoming ILMT quasar survey}
	4939 confirmed quasars from SDSS (Sloan Digital Sky Survey) DR16Q \citep{Lyke_2020} are located in the ILMT field. The SDSS catalogue was chosen over the Milliquas catalogue \citep{milliquas}, which contains twice as many quasars, because of the universal availability of magnitudes in SDSS bands and good overlap with the ILMT field of view. These can be imaged nightly (barring weather and season) in SDSS g$^\prime$, r$^\prime$, or i$^\prime$, with no competition for time with other science campaigns. At the same time, its 4-metre aperture and 102 second exposures would let it exceed the depth achieved by CRTS and ZTF. According to estimates of ILMT limiting magnitude in Table\,\ref{table:lim_mag}, 95\% of the relevant quasars can be imaged by the ILMT at a signal-to-noise ratio (SNR) of 10 or greater, and half at 20 or greater. Such precision allows for finer measurement of quasar variability.
	
\begin{table}
\centering
\begin{minipage}{92mm}
\caption{Limiting magnitude of the ILMT estimated in each photometric band for a point source in an aperture of 2.5$''$ radius (conservative estimate).}
\end{minipage}
\bigskip

\begin{tabular}{ p{3cm} p{1.5cm} p{1.5cm} p{1.5cm} }
\hline
\multicolumn{4}{c}{\textbf{Estimated ILMT Limiting Magnitudes}} \\
\textbf{SNR} & \textbf{g$^\prime$} & \textbf{r$^\prime$} & \textbf{i$^\prime$} \\
\hline
10							& 22.201 	& 21.833 	& 21.337\\
20							& 21.436 	& 21.076 	& 20.582\\
\hline
\end{tabular}
\label{table:lim_mag}
\end{table}
	
\begin{figure}[t]
\centering
\includegraphics[width=\textwidth]{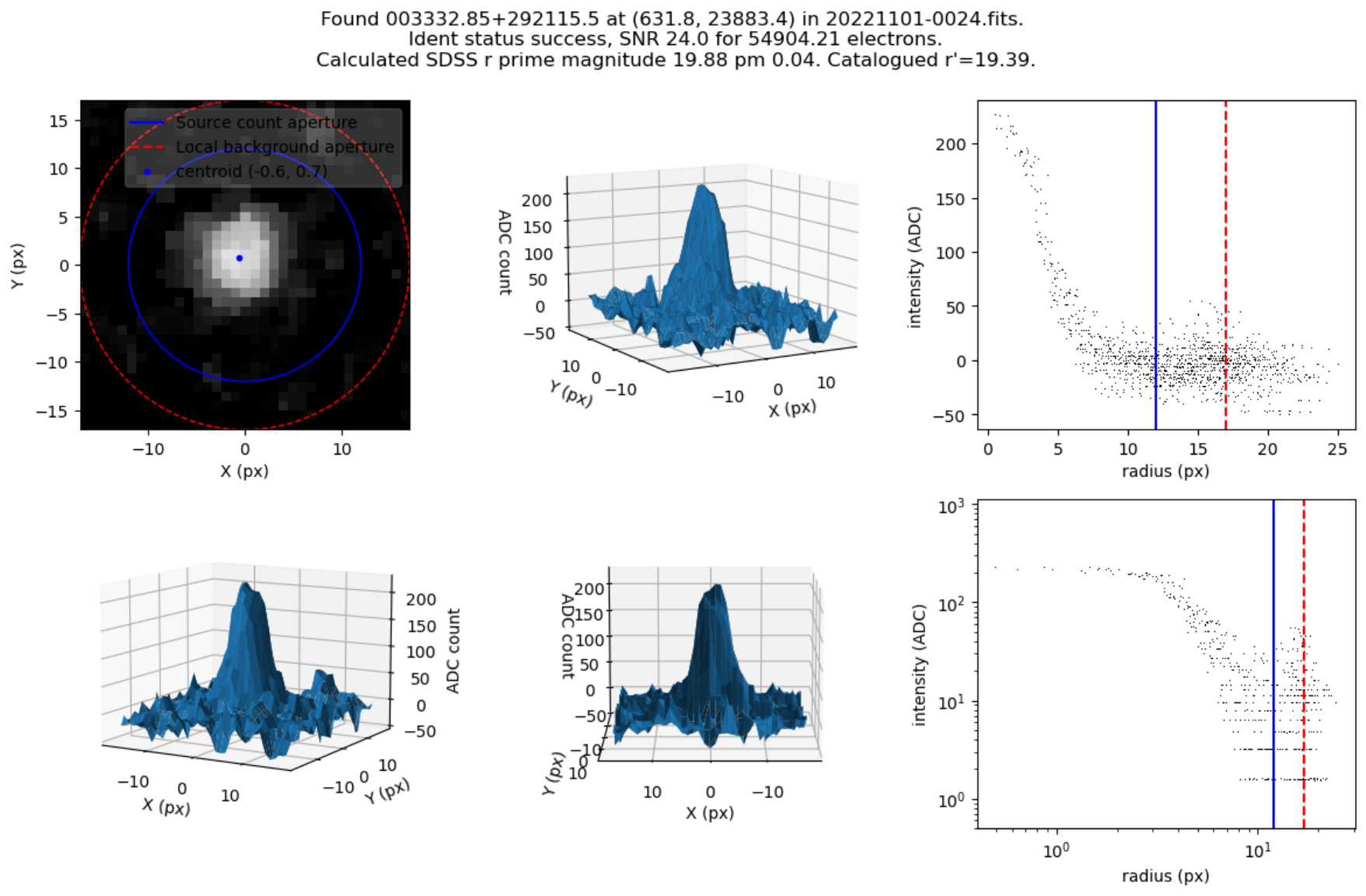}
\bigskip
\begin{minipage}{12cm}
\caption{Sample of SunPhot photometry panel with various graphical representations of the source PSF, and measured source characteristics in text form. Subject is quasar J003332.85+292115.5 in SDSS DR16Q, as imaged by ILMT.}
\label{fig:sample_aphot}
\end{minipage}
\end{figure}
	
\subsection{SunPhot}
	
	SunPhot (rhymes with "sunspot") is a Python photometric data pipeline being developed at UBC to extract light curves directly from calibrated ILMT images and make them available for study. The source code, as well as detailed documentation of version 2.1 which is current at the time of writing \citep{Sun_Thesis}, are available from the corresponding author upon request.
	
\begin{figure}[t]
\centering
\includegraphics[width=\textwidth]{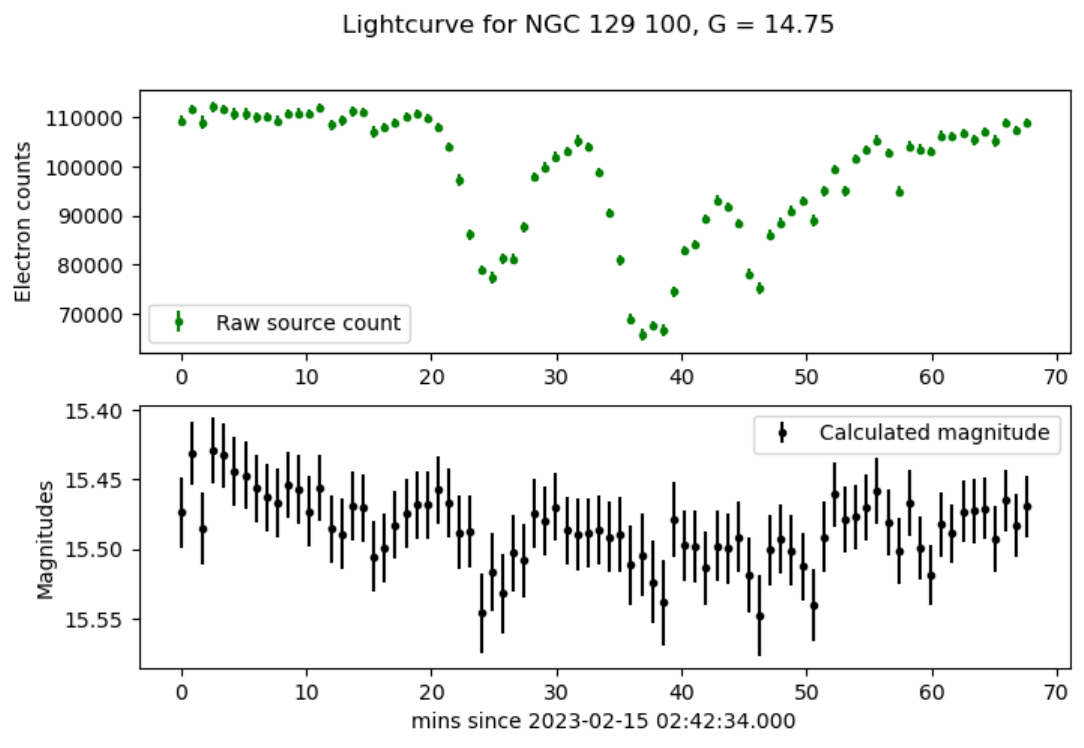}
\bigskip
\begin{minipage}{12cm}
\caption{Sample of SunPhot lightcurve for the star NGC 129 100 using data from the Plaskett telescope (not ILMT). The top panel is the raw ADU count, and bottom panel shows calculated r$^\prime$ magnitudes using OCS for photometric calibration. The variations in the electron counts are caused by cloud passages and are mostly mitigated once converted to magnitudes. The star is not known to be variable.}
\label{fig:sample_lightcurve}
\end{minipage}
\end{figure}

	SunPhot accepts images that have been astrometrically and photometrically calibrated (the OCS software package \citep{Hickson_2019} is recommended for this purpose), together with a catalogue of sources of interest. The light curve of each successfully measured source is output in CSV table format and PNG image format (Fig.\,\ref{fig:sample_lightcurve}). The lightcurve contains imaging times, flux in linear units and magnitudes with uncertainties, SNR, imaging band, and flags for measurement irregularities. 
	
	A graphical photometry panel (Fig.\,\ref{fig:sample_aphot}) can be produced for each measurement if human review is necessary, but this is very resource-intensive. The text at top contains the catalogued name of the source, pixel coordinates and image where the source was found, measurement status, SNR, measured flux in electrons and SDSS magnitudes, and the catalogued magnitude of the object if available. The six plots are various representations of the observed point spread function (PSF) of the source. If the PSF was successfully fitted, the fit and parameters will also be shown.

\section{Discussion}
	As the image quality of the ILMT improves, development of SunPhot will focus on reducing aperture sizes and implementing photometric calibration and PSF measurement. When the image PSF is well characterised, the aperture does not need to include all of the flux from the source, as the missed flux can be extrapolated with a higher overall SNR. A mathematical description of the PSF also allows flux measurements using the PSF fit method, which is generally more precise than aperture photometry.
	
	A challenge of performing photometric calibration on TDI images is that different parts of the image are captured at different times since the RA direction also functions as a time axis. In less-than-ideal atmospheric conditions, this variation dominates the uncertainty in the OCS photometric calibration. Under clear skies, the uncertainty is instead dominated by OCS's use of the Gaia catalogue for calibration. Gaia provides a conversion between Gaia magnitudes and colours and SDSS magnitudes, which is a best fit to the observed data but has significant scatter \citep{Riello_2021}. However, the exact relationships differ for different types of stars, and even more so for quasars versus stars. For ILMT surveys, a catalogue of photometric standard sources is desired. Ideally, these are sources that have a known, consistent brightness. The brightness must be known either in SDSS bands, or in Gaia bands with a known conversion for that type of star. The catalogue should contain enough sources so that multiple calibration sources can be found close enough in right ascension to any imaged quasar.
	
	As of March 2023, ILMT images taken under clear sky conditions have a FWHM between 2 and 3 arcseconds. This is primarily caused by a small wobble in the mirror rotation, which is being corrected by the technical team onsite. This effect manifests as a flattening of the PSF. From the preliminary data, an SNR of 10 is typically achieved at around magnitude 20.3 in the r$^\prime$ band, which is already on par with CRTS and ZTF abilities. This is not the result of a formal analysis. Once the PSF and aperture sizes are reduced, the ILMT appears to be on track to achieve the limiting precision in Table\,\ref{table:lim_mag}.

\begin{acknowledgments}
The 4m International Liquid Mirror Telescope (ILMT) project results from a collaboration between the Institute of Astrophysics and Geophysics (University of Li\`{e}ge, Belgium), the Universities of British Columbia, Laval, Montreal, Toronto, Victoria and York University, and Aryabhatta Research Institute of observational sciencES (ARIES, India). The authors thank Hitesh Kumar, Himanshu Rawat, Khushal Singh and other observing staff for their assistance at the 4m ILMT.  The team acknowledges the contributions of ARIES's past and present scientific, engineering and administrative members in the realisation of the ILMT project. JS wishes to thank Service Public Wallonie, F.R.S.-FNRS (Belgium) and the University of Li\`{e}ge, Belgium for funding the construction of the ILMT. PH acknowledges financial support from the Natural Sciences and Engineering Research Council of Canada, RGPIN-2019-04369. PH and JS thank ARIES for hospitality during their visits to Devasthal. B.A. acknowledges the Council of Scientific $\&$ Industrial Research (CSIR) fellowship award (09/948(0005)/2020-EMR-I) for this work. M.D. acknowledges Innovation in Science Pursuit for Inspired Research (INSPIRE) fellowship award (DST/INSPIRE Fellowship/2020/IF200251) for this work. T.A. thanks Ministry of Higher Education, Science and Innovations of Uzbekistan (grant FZ-20200929344).
\end{acknowledgments}

\begin{furtherinformation}


\begin{authorcontributions}
This work results from a long-term collaboration to which all authors have made significant contributions.
\end{authorcontributions}

\begin{conflictsofinterest}
The authors declare no conflict of interest.
\end{conflictsofinterest}

\end{furtherinformation}

\bibliographystyle{bullsrsl-en}

\bibliography{S11-P14_SunE}

\end{document}